\documentclass[twocolumn,twoside,slac_two]{revtex4}
\usepackage{graphicx}
\usepackage{fancyhdr}
\usepackage{graphics}
\usepackage{epstopdf}
\usepackage{textpos}
\def\pt{\ensuremath{p_{\mathrm{T}}}}

\begin{document}

\title{Measurement of the muon inclusive cross section in pp collisions at $\sqrt{s}$ = 7 TeV with the ATLAS detector}
\author{
\centering
\begin{center}
Silvia Franchino
\end{center}}
\affiliation{\centering Universit\`a di Pavia and INFN Pavia, Pavia, 27100, Italy}
\begin{abstract}
The measurement of the muon inclusive differential cross section d$\sigma$/dp$_{T}$ in pp collisions at $\sqrt{s}$=7 TeV with the ATLAS detector is presented. The analysis is performed in the pseudorapidity interval $|\eta|$ $<$ 2.5 for muon of transverse momentum 4 $<$ p$_{T}$ $<$ 100 GeV and with an integrated luminosity of 1.4 pb$^{-1}$.
The result is compared with the next-to-leading order with next-to-leading log high p$_{T}$ resummation prediction for the heavy flavour production and with MC@NLO prediction for W-Z bosons production. 
The measurement is sensitive for the first time to the next-to-leading log contribution to the heavy flavour production in hadronic interactions.
\end{abstract}
\maketitle
\thispagestyle{fancy}
\section{INTRODUCTION}
High transverse momentum muons are a key signature of interesting processes in hadronic interactions.  The spectrum of muons produced close to the interaction point in pp interactions is sensitive to heavy flavour production in the range 4 $<$ \pt $<$ 100 GeV.  
A measurement in this region is useful to optimise theoretical models and effective MC generators in order to improve the evaluation of the heavy flavour production mechanism.
Nowadays theoretical computations from pQCD  are available allowing predictions of the muon spectrum from the QCD lagrangian and data-driven non perturbative corrections, dependent upon fragmentation functions extracted from LEP data and heavy flavours decays and form factors extracted from B-factories. These calculations have been performed at next-to-leading order (NLO) + next-to-leading log (NLL). No experiment up to now was sensitive to the NLL contribution.
 At high p$_T$ the NLO$+$NLL prediction becomes in fact more accurate due to the reduction  of the scales uncertainty and the deviation between the NLO computation and the NLO$+$NLL becomes  larger.   Here is presented for the first time the sensitivity to  the next-to-leading log resummation at high \pt~  at an hadron collider. 
The inclusive muon yield provides also an important input to the muon trigger tuning and to the  understanding of the background sources in the search for new particles with muons in the final state.
\section{THE MEASUREMENT}
The measured differential cross-section within the kinematic acceptance of the muon is defined by
\begin{equation}
\frac{\Delta\sigma_i}{\Delta p_{{\rm {T}}_i}} =  \frac{N_{\mathrm{sig}_{\,i}}}{\Gamma_{{\rm{bin}}_{\,i}} \cdot \int{\cal L} dt} 
\cdot 
\frac{C_{\mathrm{migration}_i}}{\epsilon_{\mathrm{(reco+PID)}_i} \cdot \epsilon_{\mathrm{trigger}_i}},
\label{eq:cross_section}
\end{equation}
where 
$N_{\mathrm{sig}_{\,i}}$ is the number of signal electrons or muons with reconstructed
$p_\mathrm{T}$ in bin $i$ of width ${\Gamma_{{\rm{bin}}_{\,i}}}$,
$\int{\cal L}dt$ is the integrated luminosity, 
$\epsilon_{\mathrm{trigger}_i}$ is the trigger efficiency and
$\epsilon_{\mathrm{(reco+PID)}_i}$ is the combined reconstruction and identification efficiency.
$C_{\mathrm{migration}_i}$ is the bin migration correction factor, 
defined as the ratio of the number of  muons in bin $i$ of true \pt~ and the number in the same bin of reconstructed \pt. Each of these contributions has been measured or estimated,  as shown in \cite{emuinclus}.\\
From the measured inclusive prompt muon cross section, we subtract the contribution from $W$/$Z$/$\gamma^{\ast}$ production in order to obtain the cross-section corresponding to the  decays of heavy-flavour hadrons produced in the $pp$ collisions to  muons. As better explained in Section 2.3, this contribution is estimated from the monte carlo simulation (normalized to measurements).\\
The spectrum of muons from heavy-flavour decays is calculated in a theoretical framework, FONLL \cite{FONLL}, that allows direct comparison with the data.
FONLL is based on three main components: the heavy quark production cross-section calculated in perturbative QCD by matching  the Fixed Order NLO terms with NLL 
high-p$_\mathrm{T}$ resummation, the non-perturbative heavy-flavour fragmentation functions determined from  $e^+e^-$ collisions and extracted in the same framework, and the decays of the heavy hadrons  to muons using decay tables and form factors from $B$-factories.

\subsection{Data sample, selection efficiency and signal extraction }
The analysis is based on a data sample collected at $\sqrt{s} = 7$~TeV  during April-August 2010. The  integrated luminosity used is $1.42\pm 0.05$~pb$^{-1}$. Simulated data samples have been generated in order to estimate backgrounds  and correct for the trigger and reconstruction efficiencies and  the resolution of the detector.\\
Events were selected using the hardware-based first level muon trigger; a collision vertex with more than two associated tracks was also selected. Muon candidates were required to pass several identification criteria based on track matching and quality.\\
The trigger efficiency for the muon candidates is evaluated  using events recorded by an independent trigger  based on calorimeter information alone. \\
 The efficiency for the lower threshold trigger is found to be 68\% at \pt~$=$~4 GeV and to reach a plateau of 84\% at 9 GeV. The 10 GeV threshold trigger efficiency is constant for \pt~$ > 16$ GeV with a value of 74\%. (The muon trigger efficiency is dominated by the limited acceptance of the muon trigger chambers). \\
Reconstruction efficiency has been determined from high-statistics simulated muon samples from  heavy-flavour hadron and $W$/$Z/\gamma^{\ast}$ decays, with correction factors  being determined by comparing the simulation-derived efficiencies with those observed in data.  
The overall reconstruction efficiency is found to be 85\% at $\pt=4$~GeV, reaching 95\% at 7 GeV. The plateau value of 95\% is the same for both isolated and non-isolated muons.\\
In addition to signal muons from charm, beauty and  $W$/$Z$/$\gamma^{\ast}$ decays, the selected candidates comprise a significant fraction of background  muons from pion and kaon decays in flight and misidentified muons from  hadronic showers in the calorimeter that reach the muon spectrometer (MS) and are wrongly matched to a reconstructed inner detector (ID) track.
The muon reconstruction provides independent information on the $\pt$ of the track reconstructed in the ID  and in the MS. 
The difference in \pt , $\Delta \pt = \pt^{\rm ID} - \pt^{\rm MS}$, where both momenta are extrapolated  to the interaction point, is sensitive to the origin of the muons: signal, early-$\pi/K$, and late-$\pi/K$ or fakes. A fit to the data distribution is performed to extract the signal component using templates from the simulation.   \\
The signal purity of the sample was determined to be from  45\% at $\pt = 4$~GeV to 90\% at 40~GeV in the region of the $W$/$Z$ Jacobian peak.\\
The muon momentum resolution has been studied using tracks from the decays $Z \to \mu^{+} \mu^{-}$ and $J/\psi \to \mu^{+} \mu^{-}$.

\subsection{Inclusive muon differential cross section}
 \begin{figure*}[t]
\centering
\includegraphics[width=70mm]{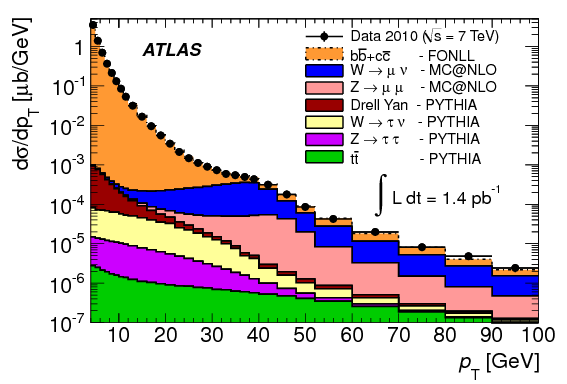}
\includegraphics[width=55mm]{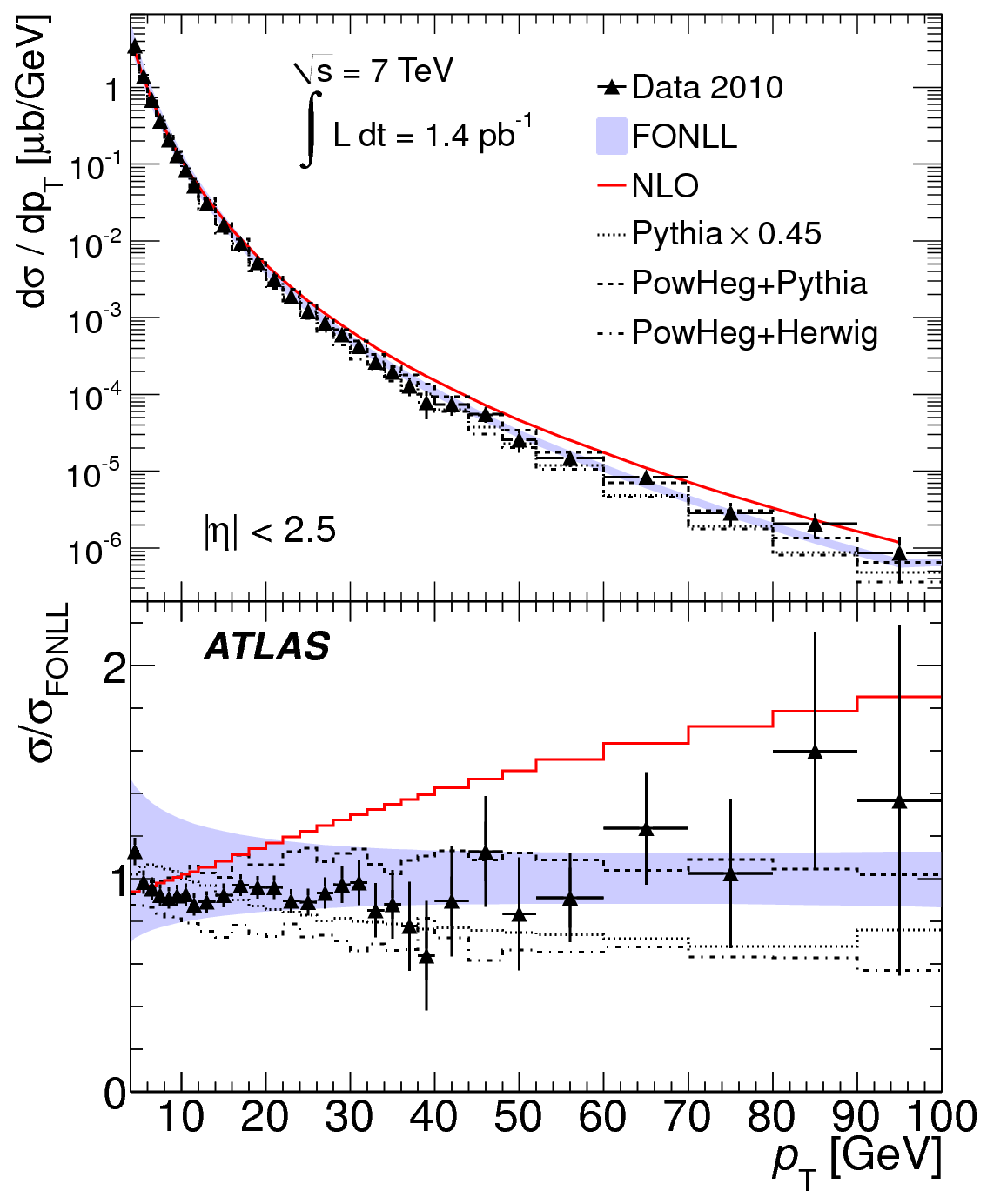}
\caption{(Left) Muon differential cross-section as a function of the muon transverse momentum for $|\eta| <$ 2.5  compared to theoretical predictions from each contributions. \\
(Right) Muon differential cross section as a function of \pt~   after subtraction of the W/Z/$\gamma^{(*)}$ contribution and compared to the prediction of the FONLL and NLO calculations and predictions of PYTHIA, POWEG+PYTHIA and POWEG+HERWIG.   The ratio of the measured cross-section and the other predicted cross-section to the FONLL calculation is given in the bottom of the plot.  } \label{mucross}
\end{figure*}
The signal fraction of the muon transverse momentum spectrum has been corrected for the trigger and reconstruction efficiencies and unfolded from the detector response. Figure~\ref{mucross}  Left shows the resulting inclusive muon differential cross-section for muons within $|\eta| < 2.5$ as a function of \pt, compared to the overall theoretical expectation.
 The  expected $W$/$Z$ component comes from {\tt MC@NLO} using the {\tt CTEQ6.6} PDFs,  normalised to the cross-section for muons measured by ATLAS~\cite{ATLAS_W_Z}. The FONLL prediction is used for the heavy-flavour component and the remaining, small contributions are obtained from {\tt PYTHIA} simulation. 
 The theoretical uncertainty is dominated by the heavy-flavour prediction, being approximately 20\%, and is not shown in the figure. The summary of systematic uncertainties on the muon cross-section measurement is shown in Table 1.\\
Integrating over the full 4-100~GeV \pt~range, in $|\eta| < 2.5$, we find a fiducial cross-section for inclusive muons of 
\begin{displaymath}
\sigma^{\mu}_{\rm Inc.} = 6.55 \pm 0.01{\rm (stat.)} \pm 0.37{\rm (syst.)} \pm 0.22{\rm (lumi.)} ~\mu {\rm b}.
\end{displaymath}
\begin{table}[h]
\caption{\it\small Summary of systematic uncertainties on the muon cross-section measurement.
The uncertainties apply in the \pt~bins of the measurement; an interval or upper limit is given where the 
uncertainty varies as a function of~\pt.}
\begin{center}
\footnotesize
\begin{tabular}{l|l}
\hline
                                         &  Cross-section \\
Source of systematic uncertainty         &  uncertainty \\
                                         & (\%)  \\
\hline 
\hline
Possible bias in signal extraction & 3 \\ 
Early-$\pi/K$ fraction & $<$2 \\
Stat. uncertainty on  signal extraction templates & 1$-$8 \\
\hline
Efficiency scale factor & 1.2 \\
\hline
Trigger efficiency  control sample statistics & 0.4$-$0.9 \\
Trigger efficiency  control sample bias & $<$2.3 \\
Trigger efficiency  background bias & 0.2$-$0.7 \\
Trigger efficiency  mis-modelling of signal fraction  & 0.5$-$0.7 \\
\hline
Unfolding procedure & 0.1$-$1.2 \\
\hline
Integrated luminosity                      & 3.4 \\
\hline
\hline
Total  &  5$-$8\\
\hline
\end{tabular}
\end{center}
\label{tab:Summary_muons}
\end{table}

\subsection{Comparison with theoretical predictions}
The measured muon cross-section compared with the predicted ones,  is given  over the full \pt~(4-100 GeV)  and pseudorapidity ($|\eta| <2.5$)  range in Fig.~\ref{mucross} (right). 
The measured heavy-flavour cross-section  is compared to the FONLL calculations, with a rigorous evaluation of the associated uncertainty shown as a band in the figure. The theoretical uncertainties  are around 20-40\%, decreasing with \pt. The dominant contribution comes from the renormalisation and factorisation scales  (up to 35\% at low \pt). The uncertainty on the heavy quark masses contributes up to 9\% at low \pt, and the PDF-related uncertainty  is below 8\% over the whole~\pt~ range. Uncertainties arising from the value of $\alpha_s$ and from the non-perturbative  fragmentation function are less than few percent. \\
The results are also compared to the NLO predictions of the {\tt POWHEG} program, interfaced to either {\tt PYTHIA} or {\tt HERWIG} for the parton shower simulation,  and to the LO plus parton shower predictions of {\tt PYTHIA}. Whereas {\tt POWHEG+PYTHIA} agrees well with the FONLL predictions,  
{\tt POWHEG+HERWIG} predicts a significantly lower total cross-section.  Less than half of this difference may be accounted for by the
different heavy-flavour hadron decay models, checked by implementing  a common decay simulation, {\tt EVTGEN} ,  for
both showering and hadronisation programs.   {\tt PYTHIA} (LO) describes the \pt-dependence well but predicts approximately a factor two higher total 
cross-section.\\
Comparisons are also made to the NLO central value expectation obtained from the FONLL program by excluding the NLL resummation part of the perturbative QCD calculation. Data deviate significantly from the NLO prediction, showing sensitivity to the NLL resummation term in the pQCD calculation for the first time in heavy-flavor production at hadron colliders.\\

\section{Conclusions}

The differential cross-section of  muons arising from heavy-flavour  production  have been measured in the  \pt~range 4 $< $ \pt~$ <$ 100 GeV within  $|\eta|<2.5$.
The theoretical predictions for heavy-flavour production from the FONLL computation are in  good agreement with the measurement. \\
Good agreement is also seen with the predictions of  {\tt POWHEG+PYTHIA}, although {\tt POWHEG+HERWIG} predicts a significantly lower total cross-section.   {\tt PYTHIA} describes the \pt-dependence well but predicts approximately a factor two higher total  cross-section.
For muons with \pt~$> 25$ GeV  a deviation from the NLO central prediction is seen, indicating sensitivity of the heavy-flavour production data to the NLL high-\pt~resummation terms. \\

%




\bigskip 
\bibliography{basename of .bib file}

\end{document}